\documentclass[prb,twocolumn, aps, amsmath, amssymb,superscriptaddress, footinbib]{revtex4-1}
\usepackage{graphicx,amsmath,amssymb}
\usepackage[usenames]{color}

\begin{document}
\newcommand{\kk}{{\bf k}}
\newcommand{\Q}{{\bf Q}}
\newcommand{\q}{{\bf q}}
\newcommand{\gk}{g_\textbf{k}}
\newcommand{\ee}{\tilde{\epsilon}^{(1)}_\textbf{k}}
\newcommand{\HH}{\mathcal{H}}
\newcommand{\oned}{quasi-1D }
\newcommand{\twod}{2D }

\title{Spin-Torque Generation in Topological-Insulator-Based  Heterostructures}
\author{Mark H. Fischer}
\affiliation{%
Department of Condensed Matter Physics, Weizmann Institute of Science, Rehovot 7610001, Israel
}
\author{Abolhassan Vaezi}
\affiliation{%
Department of Physics, Stanford University, Stanford, California 94305, USA
}
\affiliation{%
Department of Physics, Cornell University, Ithaca, New York 14853, USA
}
\author{Aurelien Manchon}
\affiliation{%
King Abdullah University of Science and Technology (KAUST), Physical Sciences and Engineering Division, Thuwal 23955-6900, Saudi Arabia
}
\author{Eun-Ah Kim}
\affiliation{%
Department of Physics, Cornell University, Ithaca, New York 14853, USA
}%

\date{\today}
\begin{abstract}
Heterostructures utilizing topological insulators exhibit a remarkable spin-torque efficiency. 
However, the exact origin of the strong torque, in particular whether it stems from the spin-momentum locking of the topological surface states or rather from spin-Hall physics of the topological-insulator bulk remains unclear. 
Here, we explore a mechanism of spin-torque generation purely based on the topological surface states.
We consider topological-insulator-based bilayers involving ferromagnetic metal (TI/FM) and magnetically doped topological insulators (TI/\lowercase{md}TI), respectively. 
By ascribing the key theoretical differences between the two setups to location and number of active surface states, we describe both setups within the same framework of spin diffusion of the non-equilibrium spin density of the topological surface states. For the TI/FM bilayer, we find large spin-torque efficiencies of roughly equal magnitude for both in-plane and out-of-plane spin torques. For the TI/\lowercase{md}TI bilayer, we elucidate the dominance of the spin-transfer-like torque. However, we cannot explain the orders of magnitude enhancement reported.
Nevertheless, our model gives an intuitive picture of spin-torque generation in topological-insulator-based bilayers and provides theoretical constraints on spin-torque generation due to topological surface states. 
\end{abstract}

\pacs{}
\maketitle

\section{Introduction} 
Harnessing the spin-momentum locking of the surface states of topological insulators holds great promise for spintronics applications. Indeed, recent experiments on TI/FM~\cite{mellnik:2014, wang:2015} and TI/mdTI heterostructures~\cite{fan:2014} observed a large spin-torque efficiency, the figure of merit for their application.
The torque measured in these two sets of experiments, however, differs quite significantly. 
While the TI/FM experiments exhibit spin-transfer- and field-like torques of comparable magnitude, the TI/mdTI has predominantly spin-transfer-like torque, and thus resembles the spin-Hall setup of heavy metal (HM)/FM bilayers.\cite{liu:2011, liu:2012, pai:2012} Its efficiency, however, exceeds the HM/FM bilayers' by several orders of magnitude.

Devices consisting of topological insulators and ferromagnetic metals have so far mainly been the focus of theoretical studies in the context of magnetotransport, where the FM affects the transport properties of the topological surface states.\cite{burkov:2010, schwab:2011, yokoyama:2014}
Most theoretical investigations of torque generation using topological insulators, however, have focused on (ideal) TI/ferromagnetic insulator (FI) hybrid structures.\cite{garate:2010, yokoyama:2010, tserkovnyak:2012, tserkovnyak:2015}
There, a current through the topological surface state mainly results in a non-equilibrium spin density due to the surface states' helical spin structure (inverse spin-galvanic effect). Adding to the Oersted field, this acts as a magnetic field on the ferromagnetic moments.\cite{garate:2010, yokoyama:2010} This effect can clearly not account for either of the two setups.

\begin{figure}[tb]
  \begin{center}
    \includegraphics{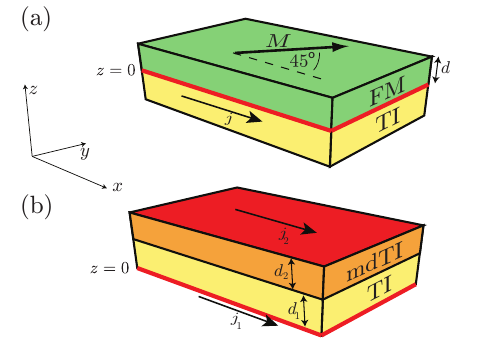}
  \end{center}
  \caption{(Color online) The heterostructures we consider in this work: (a) TI / FM bilayer~\cite{mellnik:2014,wang:2015} with a topological surface state at the inferace and (b) TI/ magnetically doped TI bilayer~\cite{fan:2014} with surface states at the two opposite surfaces (indicated in red). The current in both cases runs in $x$ direction and the in-plane magnetization $\vec{M} = M\vec{m}$ is along the in-plane diagonal.}
  \label{fig:setup}
\end{figure}

In this work, we investigate TI/FM and TI/mdTI bilayers 
assuming that in both setups the spin torque originates in the spin-momentum locking of the topological surface states. After a short description of our approach based on spin diffusion into the ferromagnetic layer,\cite{mellnik:2014} we discuss first the TI/FM bilayer with an in-plane magnetization, assuming a topological state at the interface, see Fig.~\ref{fig:setup}(a).
While it is not a priori clear that a TI next to a FM hosts a topological interface state, such a state is supported by density functional theory calculations.\footnote{Hsu et al., unpublished}
Then, we investigate the TI/mdTI structure. To describe this setup within the same scheme, we assume that both sides of the structure are `metallic', i.e., have bulk states. Furthermore, we do not expect topological interface states between the two TIs, but topological surface states on each side of the total structure,~\cite{hsu:2014} see Fig.~\ref{fig:setup}(b).
Note that while a current in the bulk may lead to additional contributions to the spin torque due to the spin Hall effect,\cite{liu:2011, liu:2012, pai:2012} we focus here entirely on the role of the topological surface states.
Finally, we discuss our findings and propose ways to disentangle the various contributions to the spin torque.

\section{Method}
The states at the surface of a topological insulator can exert a torque on an adjacent ferromagnet, which for in-plane magnetization is purely field-like.\cite{ndiaje:2015tmp} This field-like torque can intuitively be understood looking at the surface states described by the Dirac Hamiltonian
\begin{equation}
  \HH_{\kk} = v_{\rm F} (\hat{z}\times\vec{\sigma})\cdot\kk-\mu
  \label{eq:hamD}
\end{equation}
with $\vec{\sigma}$ the Pauli matrices acting in spin space and $\hat{z}$ is the unit vector in $z$ direction. Further, $\mu\neq0$ is the chemical potential away from the charge neutrality point. The velocity operator $\vec{v} = \partial_{\kk}\HH_{\kk}$
is directly proportional to the spin operator $\vec{S} = (\hbar/2) \vec{\sigma}$ and reads
\begin{equation}
  \vec{v} = \frac{2}{\hbar}  v_{\rm F}(\hat{z}\times\vec{S}).
  \label{eq:jD}
\end{equation}
While the TI has a vanishing equilibrium spin expectation, 
a finite current density 
$j_x = e n \langle v_x \rangle_{\rm neq} $ [Figs.~\ref{fig:setup}(a) and (b)], where $e$ is the electron's charge and $n$ the electron density, yields a spin density
\begin{equation}
  \langle S_y\rangle_{\rm neq} = - \frac{\hbar}{2 ev_{\rm F}} j_x.
  \label{eq:sjxD}
\end{equation}

It is important to note that in a steady-state situation of a translationally invariant system,\cite{*[{See }] [{ for effects of scattering on FM boundaries}] mahfouzi:2015tmp} which is the situation we are interested in, there is no transfer of momentum between the topological surface state and the adjacent ferromagnet. Hence, there is also no net transfer of spin from the surface states to the ferromagnet as is the case in the situation of the spin Hall effect.
However, the magnetic moments of the ferromagnetic layer couple to the surface-state spins through $\HH_{ex} = -\Delta_{ex}\vec{m}\cdot\vec{S}$ with $\vec{m}$ the magnetization direction in the ferromagnet.\cite{garate:2010, yokoyama:2010} Thus, the spin polarization on the TI surface leads to a field-like torque of the form $\vec{T} = \Delta_{ex}\vec{m}\times\langle\vec{S}\rangle_{\rm neq}$, which for an in-plane magnetization is out-of-plane.
We show in the following how for an FM layer thicker than the diffusion length, spin diffusion leads to an additional in-plane torque (Slonczewski-like torque), in a way similar to the spin-current injection in HM/FM bilayers.\cite{liu:2011, liu:2012, pai:2012}

Given the spin polarization at the TI surface, Eq.~\eqref{eq:sjxD}, as an input, we consider the diffusion of (itinerant) spins into the ferromagnetic metal and the torque they thereby exert. The diffusion (in $z$ direction) leads to a steady-state transverse spin density through~\cite{manchon:2012}
\begin{equation}
  0 = -\vec{\nabla}\cdot \vec{\mathcal{J}}_i - \frac{1}{\tau_J} (\vec{S}\times\vec{m})_i - \frac{1}{\tau_{\phi}}[\vec{m}\times(\vec{S}\times\vec{m})]_i-\frac{S_i}{\tau_{\rm sf}},
  \label{eq:spindyn}
\end{equation}
where the spin current (for the $i$th spin component) is given by
\begin{equation}
  \vec{\mathcal{J}}_i = -\mathcal{D}\vec{\nabla} S_i
  \label{eq:spincurr}
\end{equation}
with $\mathcal{D}$ the diffusion coefficient. The second term in Eq.~\eqref{eq:spindyn} describes the precession of the spins around the moments of the FM with $\tau_{J}$ the spin precession time. The third term captures the relaxation of the spin component perpendicular to $\vec{m}$ with $\tau_{\phi}$ the spin decoherence time, and the last term is the spin diffusion with time scale $\tau_{\rm sf}$. 
In the following, we use $\lambda_{\rm sf}=5$nm~\cite{bass:2007} and values for $\lambda_{\rm J}$ and $\lambda_{\phi}$ of order $1$nm ($\lambda_i^2 = \mathcal{D}\tau_i)$.
\begin{figure}[bt]
  \begin{center}
    \includegraphics{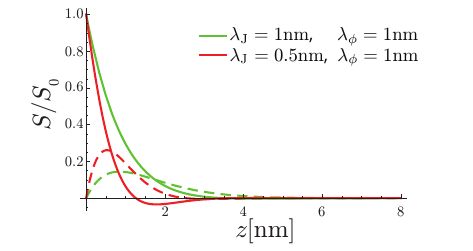}
  \end{center}
  \caption{(Color online) Spin accumulation in the ferromagnet ($d=8$nm) as a function of distance $z$ from the TI/FM boundary, where the solid (dashed) line denotes $S_\perp$ ($S_z$). For these plots, we used a spin decoherence length of $\lambda_{\phi}=1$nm and the spin diffusion length of Permalloy $\lambda_{\rm sf}=5$nm~\cite{bass:2007}. Green (red) curves correspond to a spin-precession length $\lambda_{\rm J} = 1$nm ($\lambda_{\rm J} = 0.5$nm).}
  \label{fig:spin}
\end{figure}

\section{TI/FM bilayer}
For the setup of Refs.~\onlinecite{mellnik:2014, wang:2015}, Fig.~\ref{fig:setup}(a), we solve equations \eqref{eq:spindyn} and \eqref{eq:spincurr} requiring no spin current through the outer boundary of the FM, $\mathcal{J}(d) = 0$, where $d$ is the thickness of the ferromagnetic layer. For the TI/FM interface, we assume that due to the exchange interaction, the itinerant spins of the FM right at the interface align with the spin density of the TI interface, i.e., $\vec{S}(0) = \gamma\langle \vec{S}\rangle_{\rm neq}$ with $\gamma$ of order one.\footnote{We will set in the following $\gamma =1$. Note that this choice of the boundary condition for the diffusion equation is crucial. For a spin-Hall situation, the torque is due to a spin current injected into the FM, and thus the correct boundary condition is a non-zero spin-current at the interface, i.e. $\mathcal{J}(0) \neq 0$. For realistic parameters, i.e., $\lambda_{\rm sf} \gg \lambda_{\phi}, \lambda_{\rm J}$, this results in a torque almost completely in-plane.}
With these boundary conditions, the spin distribution in $z$ direction is given by
\begin{equation}
  \hat{S}(z) = S_{\perp}(z) + i S_z(z) = S_0 \frac{\cosh[\hat{k}(z-d)]}{\cosh (\hat{k}d)}
  \label{eq:sclosed}
\end{equation}
with
\begin{equation}
  \hat{k} = \sqrt{\lambda_{\parallel}^{-2}- i \lambda_J^{-2}},
\end{equation}
and $\lambda_{\parallel}^{-2} = \lambda_{\rm sf}^{-2} + \lambda_{\phi}^{-2}$. $S_{\perp}(z)$ is the in-plane spin density and $S_0=|\vec{S}(0)\times \vec{m}|$ is the initial spin density ($z=0$), both perpendicular to $\vec{m}$.
Figure \ref{fig:spin} shows the in-plane spin density $S_\perp$ perpendicular to the magnetization (solid line) and $S_{z}$ along the $z$ axis (dashed line) for $d=8$nm.
Note that this thickness $d\approx8\rm{nm}\gg 1/k'$ with $\hat{k}=k'+ik''$. Using Eq.~\eqref{eq:sclosed}, we can thus approximate
\begin{equation}
  \hat{S}(z) \approx S_0 e^{-\hat{k}z} = S_0 \cos k''z e^{-k'z} - i S_0 \sin k'' z e^{-k' z},
\end{equation}
i.e., both components oscillate and decrease exponentially, see Figure \ref{fig:spin}. 
\begin{figure}[tb]
  \begin{center}
    \includegraphics{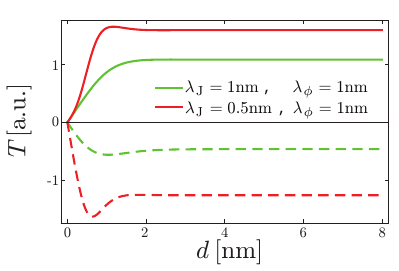}
  \end{center}
  \caption{(Color online) (a) Integrated torque as a function of the FM thickness $d$. We again set $\lambda_{\rm sf} = 5$nm, and the solid (dashed) lines denote the in-plane (out-of-plane) torque.}
  \label{fig:torque}
\end{figure}

Figure~\ref{fig:torque} shows the integrated torque as a function of the FM layer thickness $d$. Assuming the spin angular momentum to be a good quantum number, the torque is given by the spatial change of the spin current compensated by the spin relaxation,
\begin{equation}
  \hat{T} = \int_0^d dz \Big[-\partial_z\hat{\mathcal{J}}(z) - \frac{1}{\tau_{\rm sf}}\hat{S}(z)\Big],
  \label{eq:tdef}
\end{equation}
where we again use the short forms $\hat{T} = T_{\perp} + i T_z$ and $\hat{\mathcal{J}} = \mathcal{J}_{\perp} + i \mathcal{J}_z$.
Given the spin distribution in $z$ direction of Eq.~\eqref{eq:sclosed}, we find
\begin{eqnarray}
  \hat{T} &=& S_0(\frac1{\lambda_{\phi}^2} - \frac i {\lambda_{J}^2})\frac{\mathcal{D}}{\hat{k}} \frac{\sinh(\hat{k}d)}{\cosh(\hat{k}d)}\\
  &\rightarrow& S_0\frac{\mathcal{D}}{\hat{k}}(\frac1{\lambda_{\phi}^2} - \frac i {\lambda_{J}^2}).
  \label{eq:torque}
\end{eqnarray}
For the limit in the last line, we used $d\rightarrow\infty$. 
As expected from the fast decay of the spin density in Figure \ref{fig:spin}, the torque is `deposited' within only a few nanometers.
The total torque exerted on the ferromagnet as a function of the thickness $d$ thus stays constant with layer thickness. 

For the geometry described in Fig.~\ref{fig:setup}(a), the spin polarization perpendicular to the magnetization of the FM is $\sqrt{2}/2$ of the total polarization $\langle S_y\rangle_{\rm neq}$, and we find for the thick-FM limit ($d\gg1/k'$)
\begin{equation}
  \hat{T} = - \frac{\hbar}{2}\frac{\mathcal{D}}{\hat{k}}(\frac1{\lambda_{\phi}^2} - \frac i {\lambda_{J}^2})\frac{\sqrt{2}}{2}\frac{j_x}{e v_{\rm F}}.
  \label{eq:toverj}
\end{equation}
In analogy to the spin-Hall angle $\theta_{\rm SH} = (2e J_{\rm S}) / (\hbar J_{\rm C})$, which describes the spin-Hall current per charge current, we define the spin-torque efficiency
\begin{equation}
  \hat{\theta} = \frac{\hat{T}}{j_x}\frac{2e }{\hbar}=- \frac{\sqrt{2}}{2}\frac{\mathcal{D}}{v_{\rm F}\hat{k}}(\frac1{\lambda_{\phi}^2} - \frac i {\lambda_{J}^2}).
\end{equation}
For $\lambda_{\rm J}\sim\lambda_{\phi}\ll\lambda_{\rm sf}$, the out-of-plane and in-plane spin-torque efficiencies are of comparable magnitude.
Using $\lambda_{\rm J}=\lambda_{\rm \phi}=1$nm, $\lambda_{\rm sf} = 5$nm, $v_{\rm F}=5\times 10^5$ms$^{-1}$, and a typical diffusion coefficient $\mathcal{D} = 1-10$cm$^2$s$^{-1}$, we find for the in-plane and out-of-plane-torque efficiency $|\theta_{\perp}| = 0.15-1.5$ and $|\theta_z| = 0.065-0.65$. 

\begin{figure}[t]
  \centering
  \includegraphics{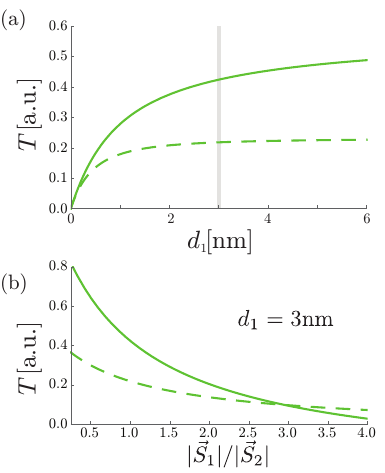}
  \caption{The two torque components as a function of the TI thickness $d_1$ for $\vec{S}_1=-\vec{S}_2$ for $\lambda_{\rm J}=\lambda_{\rm \phi}=1$nm (in the mdTI) and $\lambda_{\rm sf} = 5$nm (on both sides) and $d_2=6$nm. The solid (dashed) line denotes the in-plane (out-of-plane) torque. (b) shows the two components for fixed $d_1=3$nm [gray bar in (a)] as a function of the ratio $|\vec{S}_1|/|\vec{S}_2|$ for $|\vec{S}_1| + |\vec{S}_2|$ fixed.}
  \label{fig:doped}
\end{figure}

\section{TI/\lowercase{md}TI bilayer}
We apply the same scheme now to investigate the setup of Ref.~\onlinecite{fan:2014}, Fig.~\ref{fig:setup}(b), namely a bilayer of a TI (thickness $d_1$) and a Cr-doped TI (thickness $d_2$). At sufficiently low temperature, the doped TI exhibits ferromagnetism due to the magnetic moments introduced by Cr doping.~\cite{haazen:2012} 
Within our approach, the key difference between the TI/\lowercase{md}TI bilayer setup and the TI/FM setup is then the spatial location of the topological surface states. Assuming no topological distinction between TI and \lowercase{md}TI, we do not anticipate a topological state at the interface. Instead, we expect two surface states, one on each naked surface [see Figure~\ref{fig:setup}(b)].
These two surfaces carry the current $\vec{j}_1$ and $\vec{j}_2$ with associated  spin-polarization $\vec{S}_1$ and $\vec{S}_2$. 
Now the boundary conditions for the spin-diffusion equation~\eqref{eq:spindyn} as stated for the TI/FM bilayer has to change. First, the spin density on the two sides are $\vec{S}(0) = \vec{S}_1$ and $\vec{S}(d_1 + d_2) = \vec{S}_2$. In addition, we require that the spin density and the spin current match at the interface, i.e. at $z=d_1$. 

Figure~\ref{fig:doped}(a) shows the integrated torque of a $6$nm thick mdTI as a function of $d_1$ for $j_1=j_2$ and thus $\vec{S}_1 = -\vec{S}_2$, where we use again $\lambda_{\rm J}=\lambda_{\rm \phi}=1$nm (in the mdTI) and $\lambda_{\rm sf} = 5$nm. For $d_1=0$, i.e., no TI next to the mdTI, the contributions from the two surface states exactly cancel and upon increasing $d_1$ the torque grows monotonically with the field-like torque always smaller than the transfer-like torque. 
The two currents will in general not be identical, and Fig.~\ref{fig:doped}(b) shows the two torques for $d_1=3$nm and $d_2=6$nm, the dimensions of the experimental setup, for different ratios of $|\vec{S}_1|/|\vec{S}_2|$. As long as $|\vec{S}_1| \approx |\vec{S}_2|$, the spin-transfer-like torque dominates, in accordance with the experimental results of Ref.~\onlinecite{fan:2014}. 

\section{Discussion and Conclusions}
In this work, we analyzed the spin-torque generation in TI-based heterostructures arising from the spin-momentum locking of the topological surface states.
Considering itinerant spins that diffuse in the ferromagnetic side (either FM or mdTI), we find both an out-of-plane (field-like) and an  in-plane (Slonczewski-like) torque.
For realistic parameters, a spin-torque efficiency of the order of $|\theta|\approx0.1-1$ should be expected.
This agrees with the reported values in Refs.~\onlinecite{mellnik:2014, wang:2015} and is comparable to or larger than the largest value of spin-torque efficiency observed in HM/FM structures to date.\cite{liu:2011, liu:2012, pai:2012, miron:2011} However, we do not find as large a spin-torque efficiency as reported in Ref.~\onlinecite{fan:2014} within our approach. 

Within our model, both components of the torque stem from the combination of 
the inverse spin-galvanic effect of the TI surface and spin diffusion into the FM. 
The two torque components not only differ in their direction, but also in their behavior under $\vec{M}\mapsto - \vec{M}$: While the field-like torque changes sign, the Slonczewski-like torque does not. This can help distinguish in-plane torque arising from out-of-plane spin polarization~\cite{wang:2014b} from Slonczewski-like torque.
For `metallic' TIs, an additional spin-transfer-like torque arises from the bulk spin Hall effect. 
As transport is dominated by the surface states for thin TIs,\cite{bansal:2012} we still expect the two components of the torque to be of comparable magnitude. 
In the case of the TI/mdTI heterostructure, the fact that the transfer-like torque is more than an order of magnitude larger than the field-like torque, however, hints at a dominant contribution from the bulk.

In closing we comment on limits of the applicability of our approach to extremely thin FM layers.
As the total spin torque stays constant independent of FM layer thickness for $d\gtrsim2$nm, thin FM layers are preferable for device applications. However, our calculation treating the FM layer in $z$ direction to be in the diffusive regime relies on a FM layer that is thicker than its mean free path. For a device with an FM layer thinner than the diffusion length, the device should be modeled using a semiclassical Boltzmann approach or through quantum tunneling of spins.~\cite{xiao:2007,haney:2013, yokoyama:2014, chen:2015}. 
Our simple model can already guide ferromagnetic resonance measurements,  which do not require such thin FM layers, and help distinguish the various contributions to the spin-torque in TI based heterostructures.

\begin{acknowledgements}
The authors are grateful to Alex Mellnik and Dan Ralph for helpful discussions. MHF and E-AK acknowledge support from NSF grant no.~DMR-0955822 and from NSF grant no.~DMR-1120296 to the Cornell Center for Materials Research. MHF further acknowledges the Swiss Society of Friends of the Weizmann Institute of Science. AM was supported by the King Abdullah University of Science and Technology (KAUST). 
\end{acknowledgements}
\bibliography{ref}
\end{document}